Title: Fine structures in the optical absorption spectra of photochemical silver in silver halides? A call for further research
Author: Mladen Georgiev (Institute of Solid State Physics, Bulgarian Academy of Sciences, 1784 Sofia, Bulgaria)
Comments: 20 pages including 8 (15) figures, all pdf format
Subj-class: physics

A survey is presented of the work done so far to check earlier claims that a fine structure may be observed to occur under certain circumstances in the impurity spectral range of the optical absorption spectra of silver halides following photo stimulation in the intrinsic range. This structure, associated with the photochemical formation of silver specks, has been questioned over the years. Nevertheless, both earlier and recent findings seem to suggest the possibility that a silver fine structure may form and be observed, occasionally though due to the rapid competition of rival processes. We weigh carefully the experimental evidence on silver halides against a background of similar data on alkali halides, where competing processes run slower. We come to the conclusion that present day advances in experimental techniques may be quite adequate for providing a solid experimental basis to solve the problem unambiguously.

1. Rationale

Undoubtedly, one of the most appealing challenges for experimental solid state physics and chemistry of photosensitive materials, such as the metal halides, comes from observing the aggregation process of the metallic component. Examples are provided by the photolytic decomposition into silver clusters and halogen clusters in AgBr [1], the F-center aggregation leading to the appearance of potassium metal clusters in KCl:K [2], and the photo stimulated growth of cadmium metal clusters out of the initial metal atom phase dissolved in CdS:Cu:Cd [3]. Optical absorption changes accompanying the segregation in each of the above three cases have been employed for optical recording of information, from the photographic process in silver halides to the hologram recording in cadmium sulfide [4,5].

Now, a problem arises since the aggregation process being relatively slow at ambient temperature in the alkali halide and the cadmium sulfide alike, it seems to be very fast in the silver halide to make it too difficult to observe experimentally. Indeed, illuminating with F-light an initially disperse system of isolated F centers in the alkali halide lattice, a condensation into loosely bound F center clusters comes first, followed by the occurrence of aggregates of two, three and possibly four F centers, distinct by their optical absorption bands, as shown in Figure 1(b). What is even more important, the formation kinetics of the individual aggregate can be traced to within a reasonable certainty, as shown next in Figure 1(c). Analogous absorption bands and growth curves for CdS:Cd are displayed in Figures 2(a) and 2(b) [6]. Figures 1(a)-1(d) familiar from textbooks leave no doubt that modern spectroscopic means have the capacity of observing the details of F-center aggregation in the alkali halides.

Figures 1(b) and 1(c) have often been outlined as a motivation for the quest of analogous absorption changes accompanying the formation of latent or photolytic silver image in single crystal silver halides [7]. However, no such changes have so far been identified beyond any reasonable doubt in spite of a few affirmative assertions. "*Do not pursue this issue or you will be wasting your time*" was Professor Stasiw's unambiguous advice which he used to give me time and again during my visits to East Berlin in the late sixties of last century. Perhaps in an effort to comprehend the failure of detecting fine structure changes in pure imperfect crystals, he set forth an extensive program for investigating impurity-doped single crystals [8]. Over the years I reluctantly followed his advice to stay aside but now with the advent of femtosecond lasers and the incredible time resolution it seems that time may have come to change the attitude.

At this early stage of the paper, some personal recollections may prove stimulating. Indeed, the fine structure (FS) issue being so attractive it prompted plans for a joint research by J. Malinowski in Sofia and R. Norish in England using Professor Norish's flash spectroscopy techniques. Eventually, this plan has never been materialized. Instead, a similar method has been developed elsewhere in Sofia by replacing Norish's photo plaque detection by an electronic photo detection for monitoring the absorption change at a given wavelength and displaying it on a fast storage oscilloscope. The method was fist tested on colored alkali halides where progenitor F centers were photo ionized and the electrons trapped at other F centers to form F' centers transient at room temperature [9]. The time resolution was normally in the sub-ms range. For monitoring the entire absorption range, a special rapid scanning was devised confirming the identity of the absorbing centers. Occasional experiments were ultimately made on two silver halide samples, pure and doped with divalent cations. The resulting "sporadic" observations revealed the photo stimulation of absorbing centers in doped AgCl in the impurity dominated range near 750 nm. Unfortunately, they were not performed systematically due to the lack of funds.

In what follows we analyze the situation and suggest further experiments. It has been stated frequently that the high mobility of the silver ion makes the transition from atomic to aggregated state too fast to be followed by conventional spectroscopic means. Considering Maxwell's relaxation time, it would take only 10 μs for an interstitial silver ion to compensate the charge at a trapped photoelectron site in a Gurney-Mott ionic step [8]. The analogous time for an anion vacancy to compensate for a charge trapped at an F center in a Luty ionic step in alkali halides is much larger, the order of ~ 10 minutes [9].

## 2. Lipmann emulsion

Rather unexpectedly perhaps, the observation a photolytic silver based absorption fine structure has been claimed by Kirillov for another AgX medium used in photography, the fine-grained Lipmann emulsion [10]. Figures 3(a)-3(b) show a typical result of Kirillov's group obtained as the AgX sample has been weakly colored photo chemically by UV light. To observe a fine structure, they developed a highly sensitive differential photo-electro-optic setup in which the absorption of a colored spot of the specimen is measured relative to a reference beam passing through an uncolored spot of the specimen. Fine structures like these with the same peak positions have been reported for related systems, such as polycrystalline silver halide layers colored additively in silver vapor and silver layers evaporated *in vacuo* on a quartz substrate, as well as for gels and layers of other materials [10,11]. It is worth noting that the fine structure survived fixation which removed the silver halide or heating of the evaporated silver layers. The authors have interpreted the structure as due to the absorption of silver atoms or small groups of silver atoms.

One could argue that the structure is a manifestation of the usual scatter of experimental points but the authors stress that while the observed effect amounts to some tenths of a percent, sometimes a full percent, the mean square error is hundredths of a percent. Another argument is that the fine structure arises due to small differences of the sample thickness or layer density between the colored and uncolored spots.

The reported fine structure for an AgX material was so provocative that Frank Moser from Eastman Kodak repeated Kirillov's experiment by reproducing in great detail the fine-grained medium and the spectroscopic resolution [12]. Moser's result shows clearly that he has only observed the envelope of the photolytic silver absorption curve without being able to resolve any fine structure beneath. It has been stressed by Kirillov and Chibissov, however, that although Moser's setup has had a high integral sensitivity, it lacked the differential sensitivity required for observing the meager fine structure originating from highly dispersed silver particles.

Kirilov's claim has also been questioned by Kartuzhanskii [13]. He pointed out three points of disagreement with Kirillov's experiment: (i) The spectral measurements have been taken at a step of 5 nm, while the spectral width has been from 2 to 5 nm and the reported fine structure peaks are spaced at 10 nm on the average. (ii) The peak positions are the same whether the fine structure material is silver, gold or platinum which suggests a nature unrelated to the substance. (iii) A few works have been reported with no fine structure observed in spite of a much higher sensitivity of setup.

Although Kirillov replied, the impression has been produced that the fine structure may have been a "ghost image" rather than a real stuff possibly related to interference bands rather than to photochemistry. Anyway, this view has not been shared by Chibissov [14], who based his thermodynamic theory of latent image formation on Kirillov's fine structure [15].

### 3. Photoemission from silver into silver halide

Several years after Kirilov's original announcement a sound support has come from v. Hippel's laboratory. Gilleo has reported observing a fine structure in the photoemission spectrum of Ag into AgCl, as shown in Figure 4 [16]. The corresponding optical absorption has been too small to measure directly. The photoemission yield was detected by a method for measuring small photocurrents. The photoemission spectrum of silver evaporated onto silver chloride consisted before all of a large band well correlated with the intrinsic absorption edge of the material. This band appeared at 390 nm due to the interplay between the increasing photoconductive yield due to the increasing absorption and the increasing photo carrier density in the absorption layer resulting in an increased recombination rate. In addition to the intrinsic band, seven new bands of photoemission sensitivity have been found in AgCl having a half width of ~ 0.2 eV at 90 K which are bleachable by irradiation. These bands at 440, 500, 570, and 660 nm appeared together with silver negative, while bands at 750 and 1100 nm also seemed to be present in some cases. For silver positive, photosensitivity bands at 480 and 920 nm appeared most prominent. Except for the latter two bands, all the former bands increased in intensity with the increasing supply of electrons from the photoelectric emission. The bands at 480 and 920 nm are stimulated by electron deficiency and may thus be related to trapped holes. Photosensitivity in all the bands can be created by irradiation in the intrinsic absorption edge of AgCl at low temperature. Some of the photosensitivity bands, those at 440, 500, 570, and 660 nm appeared in AgBr as well.

The photosensitivity bands in AgCl:Ag are reported similar as far as their $v_{max}d^n$ values are concerned to Petroff's bands in KCl:K, as follows: F (440), A (480), B (???), C (660), D (570), E (500), and G (750). However, the present day assignment of the alkali halide bands does not guarantee a one-to-one correspondence with the photosensitivity bands in AgCl. Yet, one wonders if Gilleo's optical bands at 440 nm and 480 nm are not related to the optical absorptions near 474 nm attributed to P centers (positively charged silver specks) and R centers (neutral silver specks), respectively. One way or the other, Gilleo's fine structure bands are attributed to photoemission processes leading to a stepwise reduction in size of the absorbing centers and, therefore, to their inter conversion. Indeed, on losing an electron a fine-structure center will be charged positively and its charge balance restored by the detachment of one $Ag_i^+$ ion. Ultimately, a one-way repetition of the photo emission & photo detachment steps may lead to the complete deactivation of the absorbing center as in the Hershel effect. However, it should be borne in mind that Gilleo's photosensitivity bands fall within the impurity optical absorption range and may well be related to foreign atoms or molecules too.

The photoemission study being very sensitive, yet it is tempting to look for a relationship between Kirillov's and Gilleo's bands. We find some correspondence at ~ 480, 500, 570, and 750 nm but the density of Kirillov's peaks along the wavelength axis is too high to guarantee a definite result. Therefore we see Kirillov's work merits in raising the issue of a fine structure for silver halides which is substantiated by Gilleo's photoemission findings.

Nevertheless, fine structure bands very similar to Kirillov's have been reported for thin evaporated silver layers, silver gels, etc. which indicates that the fine structure, if any, has been very loosely bound to the substrate or the "solvent medium" [17]. A related fine structure has also been reported for the photo-conductivity spectra of faintly illuminated AgCl [18]. Further details of Kirillov's work can be found in his book and related references [17,18]. In addition to the experimental effort, an essential theoretical work has also been published on the optical absorption of small silver particles, later termed meso- and nano-particles [19]. The theory has helped predict the spectral range of the photochemical absorption peaks as such but has not given a clue to the conditions under which they could be observed experimentally.

## 4. Silver nanoparticles

An extensive experimental work has been carried out subsequently on the optical spectra of silver nanoparticles formed under various conditions. One of the methods used is through a collateral evaporation upon a cold substrate of [20] metal vapor and inert gas. The role of the inert component is to slow down the metal aggregation process. Figure 5 shows the absorption of tiny silver particles, $Ag_N$ aggregates (N = 1÷6), in a crypton matrix. In the particular case the spectra have been obtained at: substrate temperature 29 K, gas flow velocity $10^{17}$ $cm^{-2}s^{-1}$, metal concentration 0.8 %, precipitation time 30 s. The formation of consecutive silver aggregates $Ag_N$ at increasing N has been controlled through gradually increasing the metal concentration. However, when that concentration exceeded 1 % there has seemingly been a rapid growth to larger particles containing about 100 atoms showing evidence for the excitation of surface plasmons on small particles. The growth of the colloid absorption band according to these observations has been reported subsequently. The particle growth took place in a crypton matrix again at silver atom content of 1.6, 2.1, 3, and 5.4 % as the numeration goes from 1 through 4. The high asymmetric peak at the larger densities is the plasmon band.

We note that it might be too hard though not impossible to reconcile some of the observed peak positions in Figure 5 with the fine structure peaks reported by Kirillov. Thus the highest single point jump upwards at ~ 470 nm of Kirillov's spectra in Figure 3 may be considered reproduced by a corresponding peak attributed to $Ag_5$ in Figure 5 and there may be an argument over the peak at ~ 520 nm assigned to $Ag_4$ which seems to appear in Figure 3 as well. However, there are misfits too, such as the wide dip in the aggregate spectra near 490 nm which does not match the overall rise of fine structure activity from 450 to 500 nm.

Silver clusters and nanoparticles were prepared by reduction of silver nitrate by sodium borohydride in different ambient: (i) in water, (ii) in the presence of polyacrilate ions, and (iii) in inverse micellar solutions [21]. The formation kinetics of $Ag_4^{2+}$ aggregates and $Ag_9^+$ in the presence of polyacrilate ions was observed by the stopped-flow technique and the characteristic UV absorption spectra of the stable clusters were recorded. The optical absorption spectra of the aggregated $Ag_4^{2+}$ clusters are shown in Figure 6 while the spectra of the $Ag_9^+$ clusters are depicted in Figure 7. In both cases the clusters were stabilized by polyacrilate anions. The formation of these species was critically dependent on the way of combining the reactants. Silver nanoparticles prepared in (iii) were characterized via their UV-VIS spectra, dynamic light scattering, and transmission electron microscopy. By adding aqueous sodium borohydride solution into (iii) containing silver nitrate produced the most uniform spherical silver nanoparticles. The reaction kinetics in this case was investigated through UV-VIS spectra. The generation of silver nanoparticles occurred in parallel with the disappearance of smaller clusters, characterized by a single exponential with a time constant of about 18 s, followed by aggregation to more massive particles. Finally, the optical absorption spectra of Ag nanoparticles are displayed in Figure 7(c).

Silver clusters have been a well-studied model for some time. It is believed that a better understanding of the evolution from molecular aggregates of metal atoms to the bulk metal can follow from the study of silver clusters. Another important field of research into nanophysics is nanoclusters in zeolites. Zeolites are a class of materials with well-defined channels and cavities. The following are some of the reasons for studying silver zeolites [22]:

(i) $Ag^+$ is the only noble mono charged cation that forms a mononuclear species with appreciable stability in water. No hydrolysis occurs. Of all the noble metals only $Ag^+$ can be exchanged into zeolites from an aqueous solution. Stoichiometric ion exchange, a rare occurrence for most ions, is frequently observed for $Ag^+$.

(ii) The reversible oxidation-reduction of Ag in zeolites provides a model system for studying the formation mechanism of noble metal clusters within zeolite channels and cavities. It is also an excellent model for studying the catalytic mechanism for the dehydrogenation of hydrocarbons.

(iii) Hydrated Ag zeolites are light-sensitive materials. Silver zeolites incorporated into membrane electrodes may be used to photo catalyze the splitting of water.

(iv) Various silver sodalite materials demonstrate that silver halosodalites may be used to fabricate organized assemblies of $(Ag_4X)^{3+}$ clusters. Interactions among these clusters affect the optical properties of these materials. These are thus potential candidates for light-write and light-erase materials and they may also have applications as pressure and chemical sensors.

## 5. Photochemical silver

In an interesting study, pulsed radiolysis was used to study the early steps of coalescence of silver atoms $Ag_n$ in the presence of an electron donor (SPV – sulfonato-propylviologen), produced by that same pulse [23]. The growth of $Ag_n$ and the decay of $SPV^-$ were both followed by time-resolved optical spectroscopy. The absorbance of $SPV^-$ first held constant for a certain time-lag period, which was followed by $SPV^-$ decay correlated with $Ag_n$ growth. The essential feature of the time-lag is that a critical size has to be reached by the aggregate before it is thermodynamically capable of accepting electrons from from $SPV^-$. The cluster then grows by alternate adsorption of $Ag^+$ ions and electron scavenging. The successive pseudo-first-order components of SPV- decay are interpreted as electron transfer from $SPV^-$ towards the $Ag_n$ particles which play the role of autocatalytic growth centers. Electron micrographs of the later-stage aggregates reveal substantial size changes when the reduction is mostly achieved by the donor acting as a developer. A simulation model is developed comprising coalescence reactions between donor ans aggregates as well as an autocatalytic electron transfer beyond a critical number $n_c$. It is concluded that $n_c = 4$ when SPV- is the donor. The computer simulation of silver aggregation is also shown. It is stressed that the electrochemical potential of a small particle is only a part of the potential of a bulk particle. This is an important feature to account for while considering electrochemical processes with small (nano) particles.

Silver clusters on silver halide grains play an essential role in photographic sensitivity. There are two kinds of silver clusters acting as electron and hole traps, respectively [24]. These centers proposed to form at positively charged and neutral surface sites have been named P and R centers, respectively. Investigations by Tani have helped disclose that P and R gave rise to absorption bands at 474 nm and are destroyed by holes on exposure to light.

Reduction sensitized AgBr micro crystals have been examined with low-temperature optical absorption and optically detected magnetic resonance [25]. R-sensitization produces two closely related Ag clusters active spectroscopically. One of the centers gives rise to low-temperature emission bands at 550 and 640 nm and an absorption band at 442 nm. The other center absorbs at 430 nm and acts as a surface hole trap. Theoretical studies indicate that these prominent centers are silver dimers in two different surface configurations. We note the similarities with Gilleo's photoemission bands.

## 6. Migration of primary photochemical products

In a series of experiments during the late sixties of last century we measured the migration characteristics, such as the drift mobility and the apparent diffusion coefficient of species photo generated at one of the surfaces of wedge-shaped single crystal specimens of silver bromide. Using a photographic detection method we observed the appearance of what we tentatively call "photoelectrons" and "photoholes" at the opposite face of the crystal when driven by an electric force or by concentration diffusion. For "photohole" detection we followed their bleaching of thin (mono atomic) silver layers intensified by means of photographic development [26,27]. For detecting the "photoelectrons" we deposited surface layers of an appropriate sensitizer and applied photographic development thereafter [28,29]. The method employed was measuring the migration path of the species. Much to our surprise, the diffusion constants so determined were largely inferior to estimates based on the drift mobility by several orders of magnitude, six for the "photoholes" and seven for the "photoelectrons", as seen in Figure 8 [27,29].

The obtained drift and diffusion data and their discrepancies indicated clearly that while the observed drifting carriers were the same as the photoelectrons and the photoholes generated by the illumination, the species involved in the slower diffusion motion were not. Most likely the diffusing species were kind of small polarons or bipolarons formed as the migrating electrons and holes coupled strongly to their deformed environment. We put these polaronic "photoelectrons" and "photoholes" in quotation marks. The small polaron formation merely provides a mechanism for a slower carrier transport across the sample. Inasmuch as the formation of a small polaron takes time, the polarons do not suffice to form for too short field pulses and traversal times. This explains why the carriers in a drifting experiment have been as generated by the light pulse. Unlike them, the diffusing carriers have certainly had enough time to form small polarons.

From the measured discrepancy of the hole small polarons, $10^{-7} = \exp(-2E_{LR}/\eta\omega)$ we get for the lattice relaxation energy $E_{LR} = 8\eta\omega$ ($\eta = h/2\pi$) which signifies a considerable electron-phonon coupling strength. As a matter of fact, taking $\eta\omega = 25$ meV we obtain $E_{LR} = 0.2$ eV which is quite acceptable from the viewpoint of the small polaron theory. Similarly, we find $E_{LR} = 9.2\eta\omega$ for the electron small polaron which implies an even higher lattice relaxation energy of $E_{LR} = 0.25$ eV. Given the above estimates for the lattice relaxation energies, it is not surprising that small polarons form so effectively in silver halides.

Now, inasmuch as the hole is known to self trap stably with cubic symmetry in silver chloride but not in silver bromide, the small polaron formation in AgBr requires further consideration. If the hole does not self trap stably, then it would form transient small polarons in AgBr. Once released to its conduction band, the hole would migrate over a distance, then self-trap temporarily again at another cubic site and so on. The sequence of self-trapping & untrapping steps is remindful of the multiple trapping known from semiconductor physics and tends to reduce the apparent mobility or diffusion coefficient outright, since the carrier remains less mobile for a certain period ot its lifetime.

It can be argued further that the "photoelectrons" as small polarons otherwise centered at cubic sites may incorporate interstitial silver ions in their immediate distorted environment ($C_{4v}$ symmetry) and thereby stimulate the formation of transient silver atoms $Ag_1$ or diatoms $Ag_2$. The interstitial ions are expected to provide but transient electron traps. The "photohole" small polarons are chemically reactive as bromine atoms $Br_1$ at cubic sites or bromine molecules $Br_2$ if they occur in pairs.

Another essential conclusion to be drawn from our experiments is that the photochemical products associated with the electrons and the holes are phase segregated, as observed in other materials lately. Accordingly, phase segregation appears to be the basic prerequisite for the halide to decompose photo chemically by preventing or lessening the effect of electron-hole recombination.

In a systematic study we could measure the barrier surmounted by the diffusing "photohole" [30], even though we failed to do so for the "photoelectron". The hole barrier can be made use of while addressing the nature of the diffusing "photoholes". Unfortunately no spectroscopic measurements on the primary electron or hole polarons were made to verify their exact nature. Nevertheless, our migration experiments lend support to the fine-structure hypothesis for the silver halides. Indeed, only small-size silver or halogen particles could traverse the bulk of a macro-specimen to appear on the side opposite to the illuminated face, as observed. These small particles should be observable in a time resolved spectroscopic experiment.

It would be fair to stress that alternative to the itinerant though stable small polaron is the multiple trapping model. In it, an immobile entity forms temporarily at a site which reduces the displacement distance by holding the diffusing electrons or holes trapped for some time before releasing them again for a subsequent migration step. This multiple trapped entity is indistinguishable from the small-polaron as far as the temperature dependence of the diffusion coefficient is concerned although there is a clear distinction between the two in that the latter is slowly itinerant while the trapped entity is not. Apart from this, their external appearance is similar.

Microscopically, a "photoelectron" could be involved in a multiple trapping sequence by some more or less identified traps having no obvious link with the silver specks. In a particular case, however, the trapped entity could well be an unstable silver atom forming temporarily at the trapping site, e.g. by combining photoelectrons with interstitial silver ions, as in e' + $Ag_n$ ↔ $Ag_{n+1}$. Alternatively, the small uni- or bi-polaron complexes may also have an obvious relationship with the silver specks or bromine molecules for this is what comes out to the surface for detection.

## 7. Discussing some prospects for an appropriate experiment

With the advent of more powerful and time resolved short pulsed lasers, especially ones in the picosecond and femtosecond ranges, as well as of fast photo detectors and spectral analyzers, the unambiguous detection of narrow absorption bands occurring under photo stimulation could not constitute any problem. Femtosecond pump and probe techniques have recently been applied to monitor early formed F centers during radiation induced defect formation in alkali halides [31]. Pumping to excite the sample has been done with the third harmonics of a Ti-sapphire laser at 8 eV. Probing has been made with white light generated by the 80-fs fundamental beam of the laser to measure the time-resolved absorption spectrum from 1.5 eV to 3.0 eV. There is no reason why these powerful techniques should not be applied to the search for fine structures in the silver halides. All in all, the techniques of modern facility pump and probe spectroscopy have the full capacity to scavenge the time interval following an upsurge of excitation light. Likewise the alkali halides, an experimental study on silver halides can be combined with *ab initio* calculations of adiabatic surfaces and spectra for a better understanding of the underlying processes. On the other hand, cross-sections, concentrations, and trap depths for electrons and holes in AgBr are available and discussed in the literature [32].

There are a few basic differences between alkali and silver halides. One is the wider band gap in the former which necessitates using ultraviolet light to generate electrons and holes, while in the latter the active light ranges in the blue part of the visible spectrum. For this reason an Ar-gas laser seems appropriate for pumping the silver halide. Another difference is in the higher migration barriers in the former which slows down the defect reactions to within the minutes range at room temperature, while in the latter it takes the order of microseconds to transfer an interstitial silver ion to a defect site. Nevertheless, the reaction paths are similar and involve almost identical steps. The experiment we proposed above is by no means unique, perhaps there are alternative ways to prove or disprove the FS hypothesis. However, the pump and probe technique seems to be the most direct means to arrive at a conclusion. We also believe that the same technique can be utilized for studying the photochemical aggregation bands in CdS and elsewhere too.

Acknowledgements. *In memoriam Professor Anton D. Katsev* whose talented expertise has played a major role in forming our personal understanding of the metal nanoparticles in silver halides. May God bless his soul!

The paper is based on an invited talk given by the author at the Nanoscience Society Meeting of the Bulgarian Academy of Sciences in Sofia during the fall of 1998.

In particular, I am grateful to Dr. Vesselin Krastev and Dr. Pete Sharlandjiev, as well as to Professor Katsumi Tanimura for their kindly providing bulk bibliographic data on the silver halides, silver nanoparticles and the defect reactions in alkali halides, respectively. I also appreciate receiving numerical data on the energies in silver halides from a recent PhD Thesis by Daniela Karashanova [33]. Thanks are due to Dr. Nikolai Starbov, her supervisor, for providing a copy in due course.

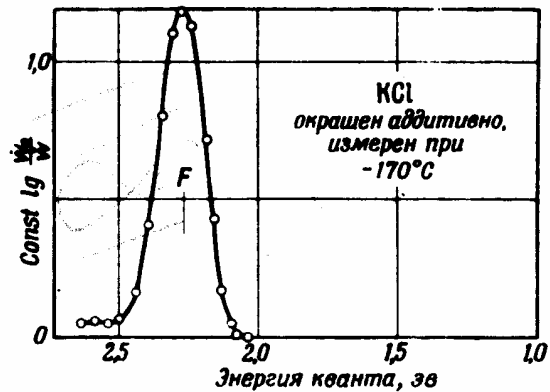

Figure 1(a): The F centered state of a KCl crystal at low temperature (-170 °C) following thermal quenching down from the temperature of additive coloration. After Ref. [7].

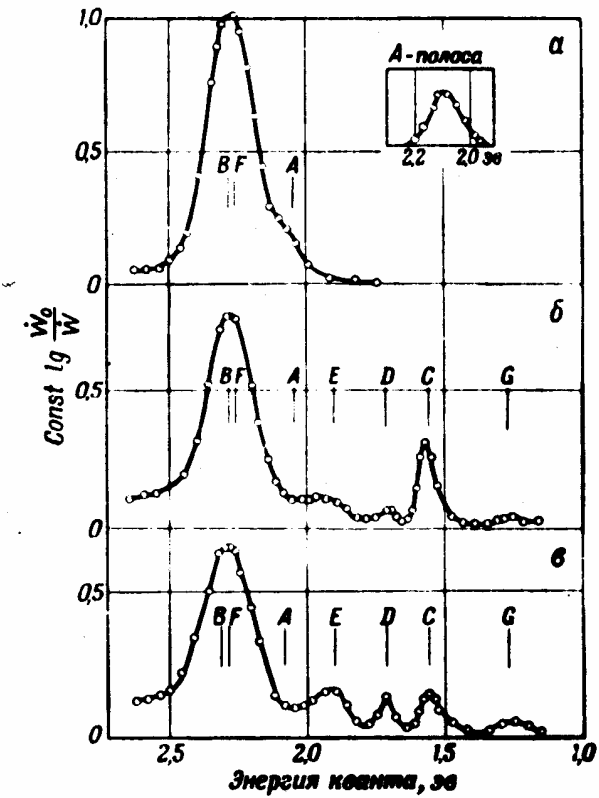

Figure 1(b): Petroff's photoaggregation bands in KCl:K (from upside down) shows the sequential development of the aggregate bands in (*a*) through (*в*) as the crystal is being illuminated within the F absorption band shown independently in (a). F is the absorption band of excess metal monomers dispersed atomic wise, A is related to F centers precipitated at $Na^+$ impurities, C is a dimer band by two F centers, D through G correspond to higher aggregates. The B band is the dual absorption of the available composite bands with optical transitions normal to the center axis. After Ref. [7].

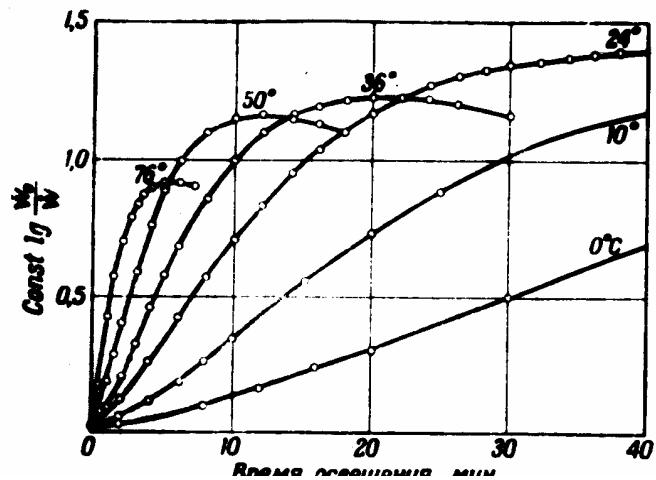

Figure 1(c): Aggregation kinetics of the C dimer band at various temperatures. The steady-state spectroscopy may be inadequate for kinetics studies at temperatures in excess of 80 °C. Yet, laser spectroscopy methods could be used beyond that temperature.

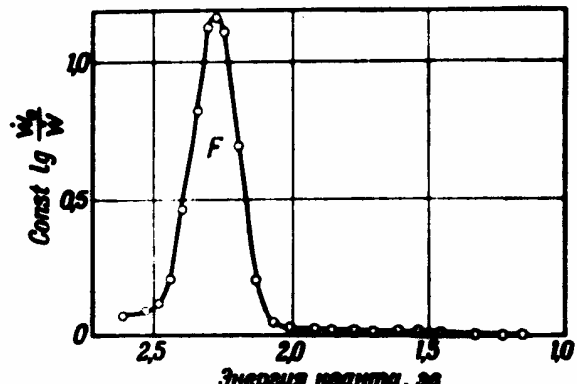

Figure 1(d): Restoration of the original F centered state of a crystal with aggregate bands after these have been deliberately bleached through illumination with infrared light.

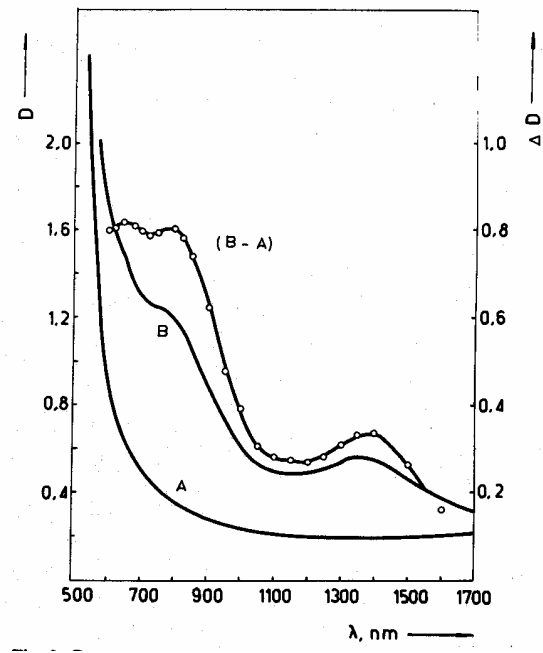

Figure 2(a): Photochemical bands in CdS single crystals doped with Cd and Cu atoms. The highest energy band at 680 nm is somewhat related to Cu, the bands at 820 nm and 1300 nm are associated with Cd aggregates. After Ref. [6].

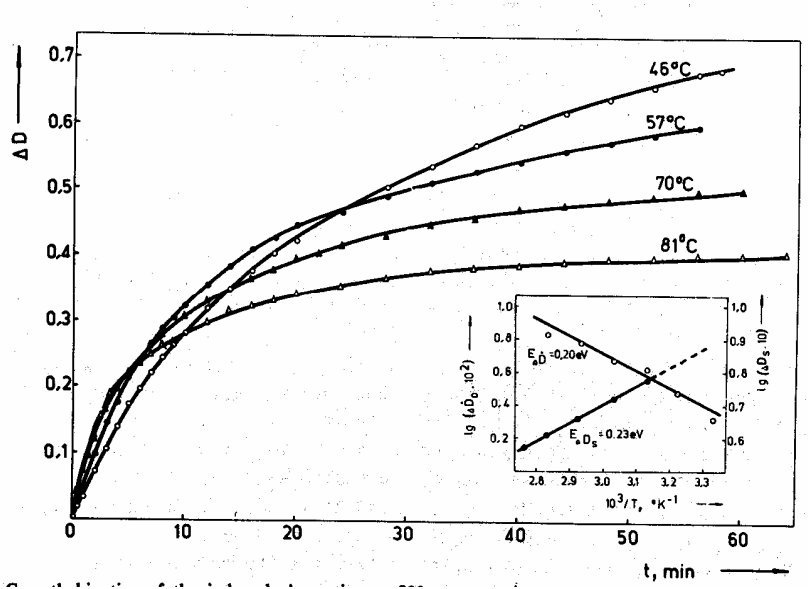

Figure 2(b): Growth kinetics of the photochemical band in CdS at various temperatures. Note the similarity with the well-known aggregate band growth kinetics in the alkali halides. After Ref. [6].

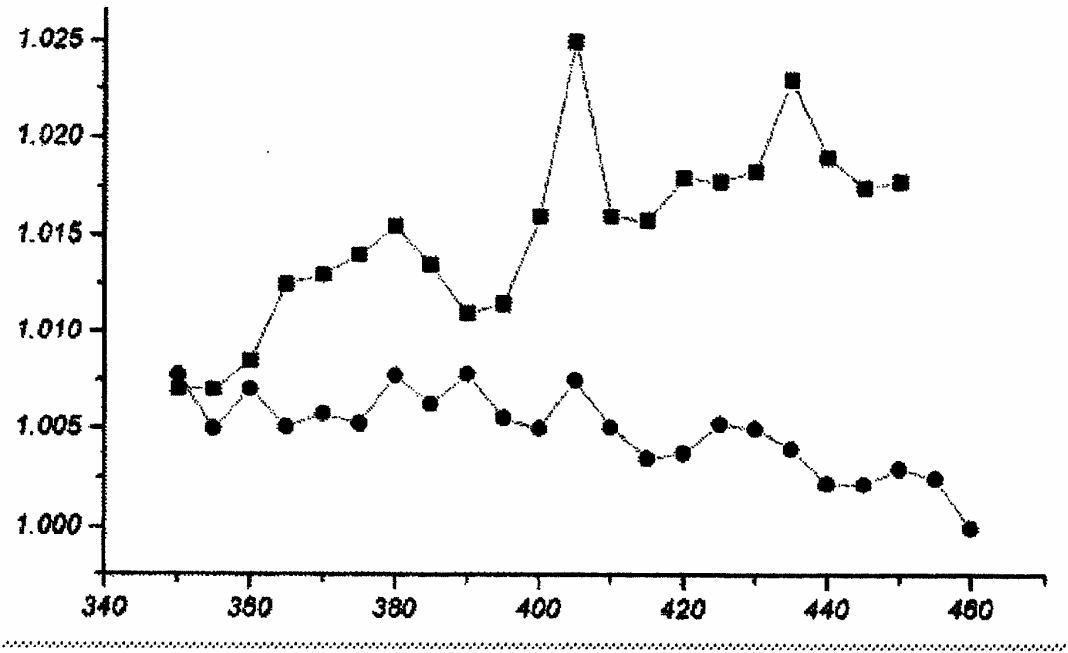

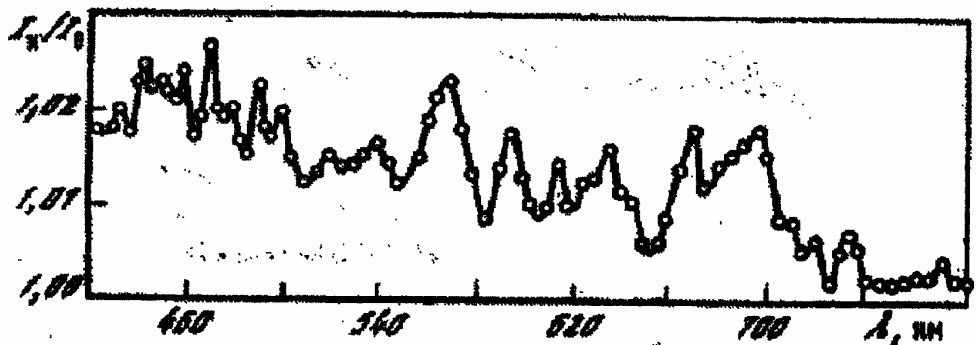

Figure 3(a) thru 3(b): Photo stimulated absorption fine structure in AgCl using data by Kirillov et al. in (a) The interconnecting lines are solely aimed at guiding the eye. Absorption fine structure of a thin film of evaporated Ag in (b). From Ref. [18].

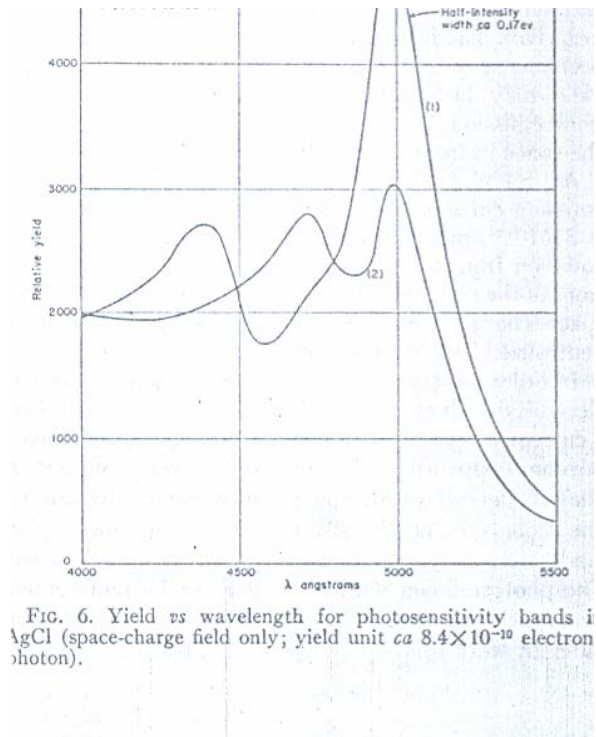

Figure 4: Photosensitive bands in AgCl formed by illumination with UV light (a). Curve (b) has been obtained by subsequent bleaching in the 500 nm band. From Ref. [16].

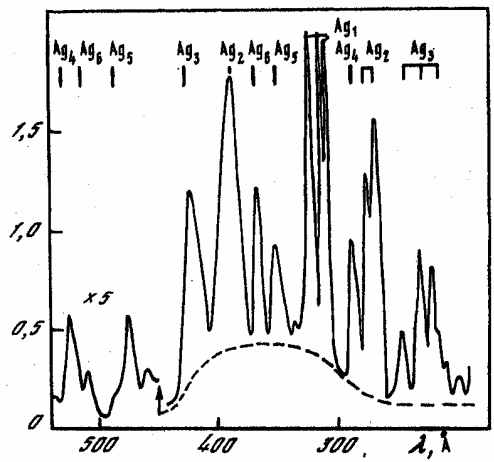

Figure 5: The optical absorption of tiny silver particles. The aggregate size is indicated on top. The envelope absorption is shown below dashed. From Ref. [20].

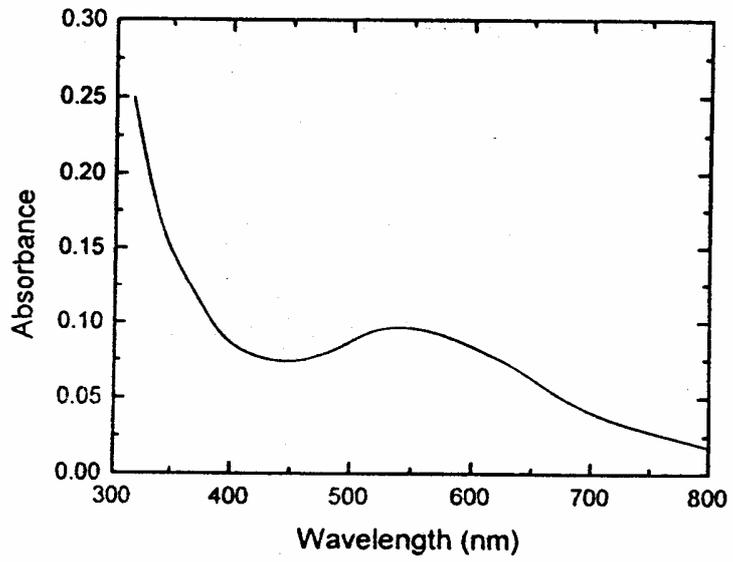

Figure 6: The optical absorption spectrum of $Ag_4^{2+}$. Silver atom quartets have been assumed to form the 'latent image specks". From Ref. [21].

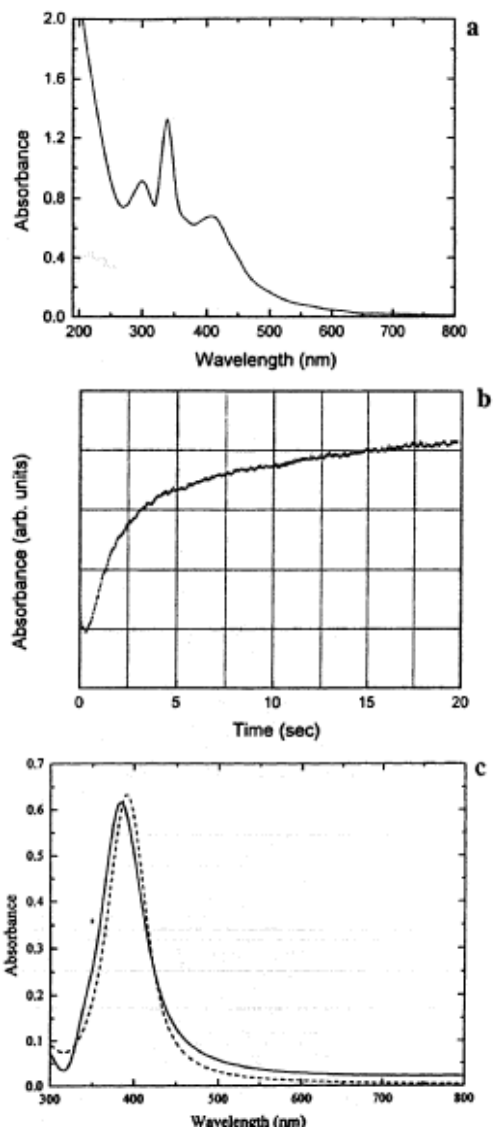

Figure 7: The optical absorption spectrum of $Ag_9^+$ in (a), the formation kinetics in (b), and the band of silver nanoparticles in (c). From Ref. [21].

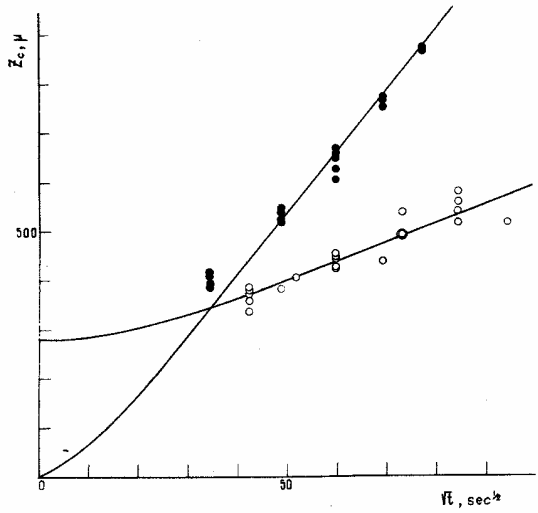

Figure 8: The diffusion paths of "photoelectrons"(hollow circles) and "photoholes" (filled circles) in silver bromide. The linear parts at longer exposure times t are proportional to the respective square root diffusion coefficients. The diffusing entities can be regarded as primary photoproducts leading to the formation of latent image specks. After Ref. [29].